# Electromigration-Aware Architecture for Modern Microprocessors

Freddy Gabbay *and* Avi Mendelson

*Abstract*— **Reliability is a fundamental requirement in microprocessors to guarantee correct execution over their lifetime. The reliability-related design rules depend on the process technology and the device operating conditions. To meet reliability requirements, advanced technologies impose challenging design rules, which have become a major burden on the VLSI implementation flow because of the severe physical constraints they impose. This paper focuses on electromigration (EM), which is one of the critical factors affecting semiconductor reliability. EM is the aging process of on-die wires and is induced by excessive current flow that can damage wires in integrated circuits. Traditionally, reliability and EM issues have been handled at the physical-design level that enforces reliability rules using worst-case scenario analysis to detect and solve violations. In this paper we offer architectural solutions that exploit architectural characteristics to reduce EM impact. The use of architectural methods can simplify EM solutions and can also be incorporated in conjunction with standard physical-design-based solutions where they offer a complementary enhancement to the current methods. Our comprehensive physical simulation results show that, with minimal area, power, and performance overhead, the proposed solution can relax EM design efforts and significantly extend microprocessor lifetime.**

## I. Introduction

CHIP reliability is an essential design requirement and is crucial to assure the correct functionality of a semiconductor integrated circuit (IC). For every product, chip vendors are required to guarantee a minimum lifetime, which depends on a reliability prediction for each chip. To meet these reliability requirements, a design-for-reliability methodology was developed that, unfortunately, is highly complicated because it depends on the expected workload, the process technology, the operating voltage, and the temperature. As part of the design-for-reliability methodology of modern processors, a workflow is defined [1,2,3] that aims to guarantee a minimum product lifetime under a specified workload (i.e., the mission profile). Given the use of new advanced process technologies and new applications such as computation-intensive infrastructures (e.g., autonomous cars, data centers and cloud computing, life-support systems, etc.), the need for high reliability has recently heightened.

The shrinking dimensions of VLSI technology, the increasing density of logical elements, and the challenging voltage and temperature operating conditions combine today to make electromigration (EM) one of the most influential factors affecting the reliability of modern systems. EM is a phenomenon related to the reliability of wires and vias in ICs. Three current models exist that are relevant for electromigration-aware design: (1) maximum [1], (2) average [2, 32] and (3) root-mean-square (RMS) currents [2]. These current models are further discussed in detail in Section 2. In this work, we focus on how the RMS current affects EM (also known as RMS-EM) within wires and vias of signals in logical cells or memory elements or that serve as interconnects between logical cells or functional units. The RMS current model is based on Joule-heating [33, 34], which is induced by alternating current. This effect leads to thermal oscillations that generate metal deformation, in turn resulting in fatigue, voids and ratcheting metal failures.

To date, the design community has focused on enhancing chip-design implementation flow [1,2,4-10] to solve EM issues, whereas few works have proposed architectural solutions. In this study, we propose a novel architecture that significantly improves reliability by reducing RMS-EM impact while relaxing the physical design efforts and significantly extending microprocessor lifetime. This study is based on the observation that numerous EM reliability concerns result from excessive write activities (or change of logical state) spread across elements of the same type (gates, logical units, or memory elements) in a nonuniform manner. This observation led us to develop enhanced resource-allocation mechanisms that uniformly distribute the write operations workload across all resources. As a result, RMS-EM hotspots induced by singular elements are minimized, and the overall IC reliability is significantly extended. Our study also enhances conventional EDA (electronic-design-automation) tools which suffer from lack of architectural information on the toggle rate of the analyzed circuit and often assume a worst-case toggling rate that may result in over design and shorter device lifetime. This work focuses on a microprocessor as a case study; however, the concepts can be applied to other ICs and applications. The contribution of this paper is summarized as follows:

1. We offer architectural solutions that exploit architectural characteristics to reduce the impact of RMS-EM.

F. Gabbay. Author is with the Electrical Engineering Department, Ruppin Academic Center, Israel (e-mail: freddyg@ruppin.ac.il).

A. Mendelson. Author is with the Computer Science and Electrical Engineering Departments, Technion – Israel Institute of Technology, Technology, Haifa, Israel 3200000 Israel (e-mail: avi.mendelson@technion.ac.il).

2. The proposed methods exploit functional building blocks characteristics such as toggle rate, hot spots and resource allocation policies.

3. The architectural method suggested herein can be incorporated in conjunction with physical-design-based solutions where it offers a complementary enhancement to the current methods.

4. The proposed solution incurs minimal cost in terms of power, performance and silicon-area overhead.

5. Our new proposed approach requires no compromise on reliability or management via the IC lifetime.

6. Our extensive experimental analysis combines architectural and EM physical simulations, which both validate the proposed architectural solution on the physical level.

The remainder of this paper is organized as follows: Section 2 introduces EM reliability challenges and reviews EM and previous works. Section 3 introduces the limitations of modern microprocessors to deal with EM, Section 4 describes the proposed EM-aware microarchitectural enhancements, and Section 5 presents both micro-architectural and physical simulation results of the proposed EM-aware microarchitecture. Finally, Section 6 summarizes the study and suggests future works.

## II. IC RELIABILITY

IC reliability has become a crucial discipline in VLSI chip design. The need for highly reliable systems has existed from the early days of computing and was mainly driven in the past by "special systems" such as mission-critical embedded systems. However, given the vulnerability of the new process technology and the appearance of new applications that require safe and reliable processing such as autonomous cars, large-scale computing-intensive systems, and life-support systems, reliability today is a fundamental requirement for most systems. The product specifications of such systems impose strict requirements on reliability through the lifetime and operating conditions. For example, the automotive industry expects an IC to function reliably for 10–15 years at a given temperature (usually about 125 °C) [11,12] and under various workloads. In data-center computing, the requirements are slightly relaxed but remain challenging: the lifetime requirement demands at least ten years, whereas the temperature can range from 105 to 110 °C with arbitrary workloads. None of these reliability-sensitive applications can afford microprocessor faults caused by reliability issues.

Over the past decade, as advanced process technologies have been introduced, the susceptibility to reliability-related issues has grown dramatically. Starting at 28 nm process technology and below, the design efforts dedicated to reliability have substantially increased. The design community has mainly tried to enhance the synthesis and place-and-route flows to handle reliability-related issues. Such flows involve substantial design efforts and, in many cases, required multiple iterations to make the IC comply with the design rules (also known as the "sign-off process"). Note that few prior studies have addressed these reliability challenges from the architecture point of view

[5-8]. The remainder of this section reviews the EM phenomenon and previous related studies.

### A. Electromigration

Electromigration (EM) is a physical phenomenon related to the reliability of wires and vias in ICs. EM causes shorts and voids in metal interconnects and decreased the median time to failure (MTF) of ICs. The occurrence of EM failure, even on a single wire, may result in overall chip failure. EM became a major concern in advanced process technologies when the geometrical dimension of wires and vias has shrunk to very small dimensions ([8]), making them highly susceptible to reliability issues. Black's equation [13] has been commonly used to model single interconnect segment median time to failure (MTF):

$$MTF = \frac{A}{J^n} e^{\frac{E_a}{K_B T}}$$

*Equation 1- EM MTF*

where $A$ is a constant, $J$ is the current density, $E_a$ is the activation energy, $n$ is a scaling factor, $K_B$ is the Boltzmann constant, and $T$ is the absolute temperature. The MTF depends exponentially on temperature; in fact, higher temperature accelerates the negative effect of EM because it weakens the atomic bonds in a wire by making them even more sensitive to EM forces. Because many new applications, and in particular control systems (e.g., in the automotive or robotics fields), are required to operate at high temperatures of 105–125 °C, this induces much greater susceptibility to EM that will be highly challenging to mitigate during IC implementation and sign-off. EM involves three electrical current models: (1) peak, (2) average and (3) root-mean-square (RMS) currents [2]. To meet the EM reliability requirements, special design-rule constraints are imposed by foundries on both peak, average and RMS currents ([14]).

When peak current is applied, even for a short duration, it induces stress through the force of conduction electrons and metal ions. When the force of conduction electrons reaches a certain strength level, it may tear atoms from the boundary of the metal and transport them in the direction of the current flow. If such current force is maintained for a long time or if current flows frequently, the wire may become malformed. Such damage to a metal wire may result in reduced wire conductivity or in the formation of voids and hillocks (i.e., short circuits) [1], all of which lead to major reliability concerns. In the peak current model, which enforces limitations on every unidirectional current flow, the current density, $J$, can be expressed as [6, 14]:

$$J = \frac{CV_{DD}}{WH} pf$$

*Equation 2 - Current Density*

where $C$ is the wire capacitance, $W$ and $H$ are the metal width and height, respectively, $V_{DD}$ is the operating voltage, $f$ is the clock frequency, and $p$ is the switching probability, also known as the toggle rate.

In the average current model, alternating current induces material backflow (i.e., reversed material flows) [2], which reduces overall material migration. This phenomenon, known as self-healing [32], is quite common in digital circuits that operate by charging and discharging metal interconnects. When the alternating current is symmetric, the impact of the average current on EM is relatively small. While EM in the peak and average current models is governed by the mobility of conduction electrons which accelerates the atomic diffusion (referred as current-induced EM), in the RMS current model [14, 33, 34], the alternating current produces thermal oscillations that deforms the metal and result in fatigue, voids, and ratcheting metal failures. This phenomenon, which is also known as the Joule-heating effect (or RMS-EM), cannot be compensated by self-healing [2]. In addition, thermal oscillations propagate to neighboring areas, with the result that nearby metals may also be degraded. RMS-EM signoff rules enforce maximum RMS current, $I_{RMS-max}$, for every net given a nominal median time to failure, $MTF_{Technology}$ (typically 10 years). Both $MTF_{Technology}$ and $I_{RMS-max}$ are specified for every process technology by the foundries ([37]). The RMS current can be relaxed if the median time to failure is compromised as indicated by Equation 3 ([14]):

$$I_{RMS-reduced} = I_{RMS-max} \sqrt{\frac{MTF_{Technology}}{MTF_{reduced}}}$$

*Equation 3- Reduced RMS current*

The MTF in the RMS current model can be calculated by the following equation ([14]):

$$MTF = \left( \left( \frac{K_1}{K_2} \right)^2 \cdot \frac{1}{C^2 V_{DD}^2} \cdot \frac{1}{F_{max} \cdot p} \right)^{\frac{n}{2}}$$

*Equation 4- RMS-EM MTF*

Where $C$ represents the capacitance load, $F_{max}$ is the maximum frequency, $K_1$ and $K_2$ are given by the following equations:

$$K_1 = A \cdot (W \cdot H)^n \cdot e^{\frac{E_a}{K_B T}}$$

*Equation 5*

$$K_2 = \sqrt{\frac{1}{t_r} + \frac{1}{t_f}}$$

*Equation 6*

Where $t_r$ is the rise time and $t_f$ is the fall time. Equation 4, which indicates that MTF is inversely proportional to the switching activity ratio, provides the motivation for our study to relax switching probability and thereby improve MTF.

Joule heating and current-induced EM have cross-coupled relations. Joule heating causes heat increase and atomic diffusion (due to temperature gradients), both result in accelerated current-induced EM rate. On the other hand, current-induced EM, increases both resistance and current density which intensify Joule heating as well due to the temperature increase. This cross-coupled positive feedback between Joule heating and current-induced EM rapidly accelerates both phenomena leading to severe reliability issues.

Handling the design rules for both maximum, average and RMS currents is highly challenging. The maximum-current constraint is mainly enforced by the physical design implementation tools that assure that the driving gates will not exceed the maximum-current limitation and by other physical design means [14]. With respect to the RMS current, the situation is more complex. Equation 4 shows that the MTF due to RMS current flow is inversely proportional to both the switching probability and the clock frequency, which means that a higher switching probability for logical elements increases the susceptibility to RMS-EM. Therefore, the MTF of wires and vias can be increased by minimizing their switching rate p. Minimizing the switching rate depends on both workload and IC architecture. In many cases, the switching probability depends on the change of logical state due to a write operation or to the use of logical elements for different computations. Read operation may also involve switching of wires state, however this usually happens on read ports shared between memory cells and therefore makes a smaller contribution to RMS-EM hotspots. Further studies on EM and its effects are available in Refs. [1,2,9,10,17,19].

This study focuses on how RMS-EM affects signal lines that are inside logical cells or memory elements or that serve as interconnects between logical cells or functional units. To relax RMS-EM impact, we propose in Section IV a novel architectural solution that exploits the relationship between RMS-EM and toggle rate.

### B. Prior Works on Electromigration

This subsection summarizes previous works on EM. The overview differentiates between works that propose EM solutions through the physical design flow and works that do so through micro-architectural or architectural solutions.

#### 1) Prior work based on physical design

EM phenomena have been broadly studied from the physical design point of view. Various studies [4,7,16] examined different interconnects such as copper or aluminum and how they are affected by EM under different process, voltage, and temperature conditions. From a physical point of view, the most common solution for EM is to widen the wires. As Equation 2 indicates, this reduces the current density and eventually decreases the effect of EM but, from the physical design viewpoint, it is not always the preferred solution because it may introduce several over-heads, such as increasing the die area, which may reduce the device frequency. In addition, a larger die may also create timing and power challenges because signals would need to travel farther.

Modern electronic-design-automation (EDA) tool vendors, in conjunction with process foundries, enforce EM-related design rules as part of the IC sign-off process. Such tools verify that interconnects and vias meet the EM design rules and identify all EM-related violations that require design fixes. EM analysis tools are even able to simulate switching activity patterns extracted from functional simulations representing real applications and take these patterns into account in the EM analysis process. When the worst-case switching patterns cannot be determined, designers often use a statistical analysis provided by the EDA sign-off tool. In this case, the design is analyzed under a given set of switching probabilities, which may lead to an over-design process. The EM sign-off process is tedious and involves many fix iterations and trials. Some of the trials involve the use of wider metals and vias and, in several cases, may even limit the clock frequency, the switching rate, and the computational workload. The combination of all these limitations may result in degraded IC performance.

A study by Dasgupta et al. [7] introduced a methodology for synthesizing the design and scheduling data transfer from the control data flow graph to the hardware buses in an EM-aware manner. Their algorithm requires that the activity be determined in advance, so it becomes tightly coupled to each specific computational use that it targets.

A broad survey of additional physical-design-based techniques to mitigate EM impact is available in Ref. [10].

### 2) Prior work based on architecture

Only a limited number of prior works have suggested architecture-based solutions to the EM problem. Srinivasan et al. [6] suggested structural duplication and graceful performance degradation techniques to handle the EM effect. Structural duplication adds spare design structures to the IC and turns them on when the original structures fail. Graceful performance degradation, however, shuts down failing structures but keeps the IC functional while degrading its performance. This approach seems to incur a major hardware overhead related to the dedicated mechanisms to detect EM degradation through normal IC operation and the need for special circuits to switch on the redundant logic. In addition, it introduces extra power and performance overhead due to the addition of redundant hardware. A similar approach to handle EM by adding redundant elements has been introduced by [38].

Abella et al. suggested [8] a novel architectural approach for "refueling" bi-directional busses by monitoring the current-flow direction each time data is transferred on the bus and suggested a mechanism that triggers current compensation whenever an imbalance occurs between the current flowing in each direction. Such a scheme could indeed relieve EM impact induced by peak current; however, it may encourage RMS-EM in the form of thermal oscillations, thereby leading to reliability concerns. In addition, given their design complexity, modern VLSI circuits do not commonly use bidirectional buses. The refueling mechanism also disrupts bus operation and may introduce a dynamic power overhead due to the reversal current.

Srinivasan et al. [5,20] suggested a dynamic reliability management approach where the processor dynamically maintains its lifetime reliability target by responding to the changing behavior of the application. This approach allows a processor with lower reliability to run correctly while compromising performance or operating conditions.

Swaminathan et al. [28] introduced BRAVO, a cycle-accurate microprocessor simulation platform to assist designers and architects to account for reliability factors. Their tool can model voltage, energy and reliability to explore the optimal operating point for applications. EM impact is modeled using analytical means (Equation 1).

Based on all this evidence, we conclude that applying only physical design-based solutions does not suffice because of the growing challenges involved by EM. The remainder of this paper describes our comprehensive architectural solution for handling RMS-EM.

## III. Distribution of RMS Electromigration Hotspots in Modern Microprocessors

Based on our previous discussion in section 2 with respect to Equation 4, our main focus in this paper is on the switching probability, p. This factor is tightly coupled to micro-architectural assumptions and application workload while all other arguments are mainly related to process technology. In addition, current RMS-EM analysis tools extract the toggle rate without detailed analysis. This may lead to over-design, and therefore our analysis becomes valuable. In our study we assume that all other factors in these equations are constant due the following reasons: The junction temperature indeed makes a major contribution to RMS-EM MTF; however, since it also depends on the workload and system cooling solution, common design flows usually consider the worst-case scenario of 105 or 125 °C in the sign-off process. As for metal width and height, the microprocessor blocks that we examine, such as ALUs, registers and memory elements, already utilize lower metal layers (typically metal 1-3), which are highly suspectable to RMS-EM. Upper metal layers are less suspectable and are mainly used for inter-block connectivity and power grid connections. We also assume operating at nominal voltage and do not assume power saving modes, such as DVS (dynamic voltage scaling), which can save power and decrease RMS-EM impact while reducing performance. Finally, the capacitance parameter depends on process intrinsic capacitance and wire length. The latter supports our interest in memories because they utilize long wires and hence are more suspectable to EM.

Since RMS-EM design rules are limited by the weakest link (i.e., the most susceptible wire), we start by examining the distribution of the switching probability over several sub-systems of a modern microprocessor which are expected to be highly susceptible to the RMS-EM effect due to hot spots caused by the toggling rate of wires. It should be noted that the EM impact on metal wires that are part of the IC power grid is out of the scope of this paper. Subsection A describes our experimental environment, and subsection B presents our comprehensive observations on RMS-EM switching probability hotspots in microprocessors.

### A. Experimental Environment

Our experiments use the sniper x86-64 simulator [21]. We modified the simulation platform and added the needed mechanisms to model the behavior and measure the characteristics required for our experiments. The simulation environment includes both a detailed cycle-level x86 core model and a memory system. Table 1 summarizes the configuration of the simulation environment (based on the Intel Gainestown core [22]). We used the simulation benchmarks Spec2017 [23,24] with *ref* inputs. The Spec2017 benchmark suite was chosen because it is provided and supported by the Standard Performance Evaluation Corporation (SPEC) and contains applications from many domains that were selected by major companies. These applications include artificial intelligence, physics, visualization, compression and document processing. In the past decades the SPEC suite has served as the de facto benchmark suite for semiconductor research and has been continuously updated by the SPEC organization to reflect the changing trends in computational applications. Every benchmark is run as a single-core workload in the main execution phase. Each experiment used 10 billion instructions.

## B. Experimental Observations of RMS-Electromigration Hotspots

This section examines switching probability hotspots which may accelerate RMS-EM in three different parts of processors microarchitecture: ALU execution units, architecture register files, and memory hierarchy sub-system. Previous studies [10, 25] support our concern that these areas involve the most intensive EM activities when running these workloads and, thus, will experience intense EM hotspots.

*Table 1- Baseline simulation model configuration*

| Core Model | |
| --- | --- |
| Frequency | 2.66 GHz |
| Execution units [time] | 3 ALUs [1 cycle] |
| | 1 FP add / sub [3 cycles] |
| | 1 FP mul /div [5/6 cycles] |
| | 1 Branch [1 cycle] |
| | 1 Load unit [1 cycle] 1 |
| | Store unit [1 cycle] |
| Pipeline | Dispatch width: 4 |
| Instruction window | 128 |
| Memory system model | |
| Block size | 64 Bytes |
| L1-D Cache | 32KB, 8-Way. |
| L1-I Cache | 32KB, 4-Way |
| L2 Cache | 256KB, 8-Way |
| L3 Cache | 8MB, 16-Way |
| D-TLB | 64 entries, 4-Way |
| I-TLB | 128 entries, 4-Way |
| S-TLB | 512 entries, 4-Way (secondary TLB) |

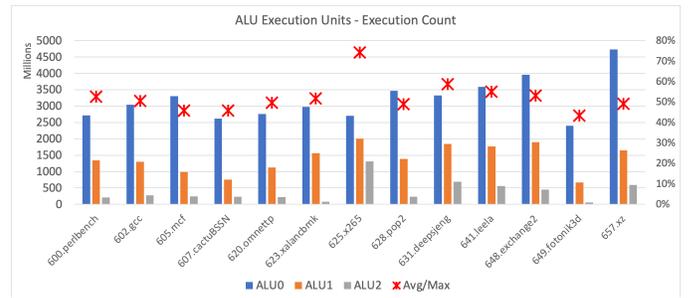

*Figure 1- Distribution of ALU execution count*

**ALUs**: Figure 1 shows the distribution of write operations among different ALUs when using the FIFO selection mechanism among all ready-to-execute instructions (all benchmarks were run for 10B instruction so the usage rate per instruction can be easily calculated). Note that ALU0 is the most-used ALU of the three available, and ALU2 is the least used, which is attributed to the fixed allocation policy of the available ALUs, whereby a higher priority is given to an ALU with a lower index. Since ALU execution time is 1 clock cycle, all ALUs become available every cycle. For example, for a program that provides exactly one instruction per cycle, we expect only ALU0 to be used. Figure 1 supports this claim and shows that ALU0 is used at over twice the rate than ALU1, and nearly ten times the rate than ALU2 for most benchmarks. In such a logical implementation, the worst-case switching factor of ALU0 dictates the worst-case RMS-EM scenario to be taken into account and applied to all ALUs.

**Register-file**: Our next set of experiments examines the switching factor on architectural registers. Figure 2 illustrates the distribution of write operations on general-purpose registers (GPRs: integer general purpose) for the Spec2017 benchmarks. The distribution clearly is not uniform; for example, the RAX register is the most-toggled register in terms of write operations, whereas the non-legacy registers are hardly used and thus are significantly less toggled than the x86 legacy registers. The root cause of these differences is the nature of compiler register-allocation algorithms. Figure 2 also shows that the ratio of the average number of write operations to the maximum number of write operations varies from nearly 7% to 33%. This measurement is another indication that the toggle rate is not equally balanced between registers; thus, the register with the greatest number of writes dictates the overall switching ratio for RMS-EM.

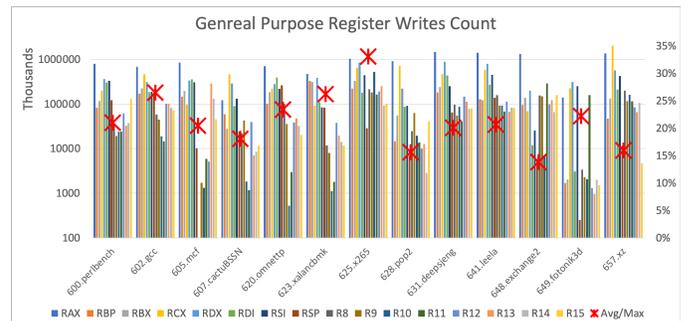

*Figure 2 - Distribution of general-purpose-register writes*

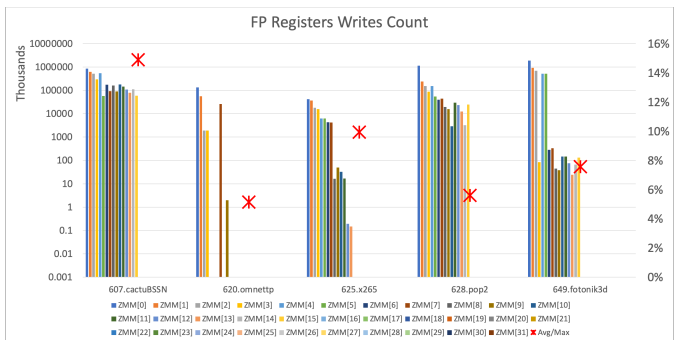

*Figure 3 - Distribution of writes to floating point registers*

Figure 3 presents the number of write operations on FP registers only for the Spec2017 benchmarks that involve FP operations. The results presented for this case are similar to the results presented in Figure 1. For FP registers, the number of writes is significantly greater in the registers with lower indexes (i.e., ZMM0, ZMM1, and ZMM2 are the registers with the highest write count). Similar to integer registers, this can also be explained by the nature of the register-allocation algorithm of common compilers. In this case, the ratio of the average number of write operations to the maximum number of write operations is even smaller, which is indicative of an even larger variance relative to integer registers.

**Memory hierarchy**: Memories are highly susceptible to EM because they employ high-density bitcells with narrow and long metal wires that toggle upon every change of logical state. SRAM memories employ lower metal layers for their interconnect, typically, metal 1 – metal 3. As opposed to upper metal layers, the width and height of lower metal layers are significantly smaller and as a result they become highly suspectable to RMS-EM. In addition, physical design tools lack the ability to handle every bitcell in an individual manner; therefore, the worst-case scenario is commonly applied to all bitcells. Since write operations are not uniformly distributed across all memory bitcells, the worst-case scenario is determined by the bitcell with the largest number of writes.

Note that the granularity of switching probability differs from one level of memory hierarchy to another; e.g., a single byte can be written in the L1 cache, but a minimum granularity of the cache line is imposed on all other levels of the cache hierarchy (assuming a line-fill mechanism). Since all bits within the write granularity have the same switching probability, we assume that they all have the same probability for failure, so conventional error-correction mechanisms may not be effective at that granularity.

We first start our analysis by examining the toggle rate of the memory hierarchy elements. Figure 4 shows ratios of the average number of write operations per memory entry. It reveals that DTLB involves significantly more write operations than ITLB. DTLB also involves nearly tenfold more write operations than STLB. A similar observation results from examining the ratio of write access of the L1-D cache to that of the L1-I cache. The L1-I cache involves write operations only upon cache line replacement, whereas L1-D maintains a much higher rate of write operations because of block replacement and each time an instruction targets a memory location. If we continue examining the write ratios of L1-D to L2 and L2 to L3 then we see that higher-level of cache memories experience a higher toggling rate.

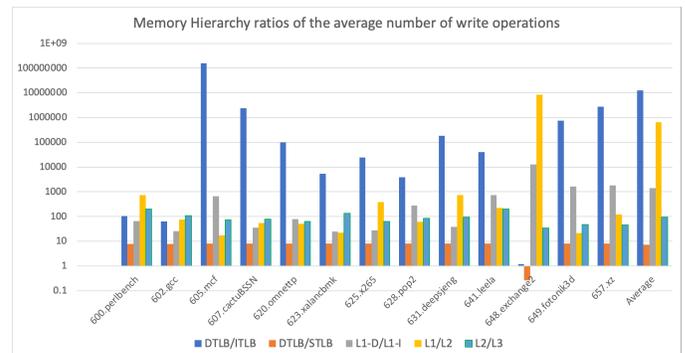

*Figure 4 - Write ratios in memory hierarchy*

Note that, although the initial observations indicate that the L1-D cache and the D-TLB have the highest write rate, we must still continue carefully watching the write distributions in the remaining memory hierarchy. In particular, it is important to monitor the write distribution to L2 and L3 cache memories. Although our experimental results show that these caches maintain lower write rates, they may be much more susceptible to RMS-EM than the L1 caches because of physical design considerations. Since both the L2 and L3 caches are significantly larger than the L1 cache, they involve higher-density memory bitcells and significantly longer and narrower interconnect metal. Equation 2 supports this argument by indicating that the current density is inversely proportional to the metal width and proportional to the wire capacitance. The interconnect metals in both the L2 and L3 caches, which use long wires, introduce a much greater interconnect capacitance than the L1 caches.

Based on this observation, the next graphs focus on how RMS-EM affects the L1-D cache, L2 cache, L3 cache, and D-TLB. In the next figures, we present histograms of write operations partitioned into five histogram bins: 0%–25%, 26%–50%, 51%–75%, 76%–90%, and 91%–100%. Each bin shows the number of cache entries with the ratio of write distributions relative to the cache entry with the maximum number of write operations. For example, 20% for bin 26%–50% means that 20% of the cache entries each experienced write operations in a ratio range of 26%–50% relative to the cache entry with the maximum number of write operations. The cache entry with the maximum number of writes is the entry that dictates the RMS-EM switching probability assumption for the entire cache. Such histograms can help illustrating the switching probability distribution among all cache entries and allow us to explore new architecture to relieve RMS-EM hotspots.

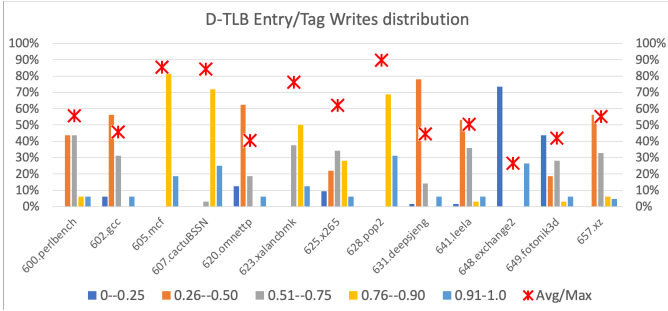

*Figure 5 - Distribution of DTLB writes*

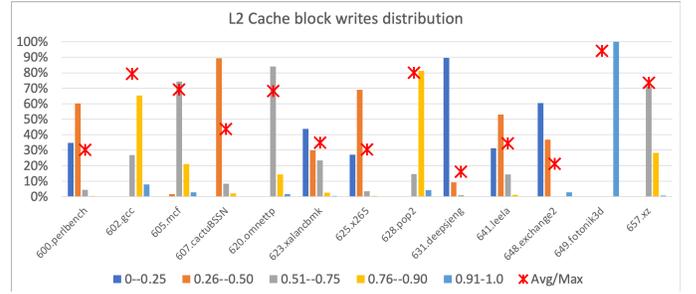

*Figure 7 - Distribution of L2 cache block writes*

Figure 5 shows the write histogram of D-TLB entries and their tags. Note that, for all benchmarks, only a small number of entries experience a large ratio (above 90% relative to the entry with the maximum number of writes); these entries dictate the overall switching rate of the D-TLB. The majority of entries experience much lower write rates. Figure 5 also presents the ratio of the average number of writes per entry to the maximum number of writes of all entries, which varies from 2% to 100%, with an average of 55%.

Figure 6 shows a histogram of writes to L1-D cache data lines. A phenomenon appears similar to that observed in the D-TLB. Only a small number of cache lines have a high write ratio (above 90% relative to the maximal data cache line), whereas the majority of cache lines experience much lower write ratios. In most of the benchmarks, the ratio of the average number to the maximum number of writes is less than 30%, whereas the average ratio is 33%.

Figure 7 shows histogram cache writes for the L2 cache data lines. The observations, in this case, are similar to those for the L1-D cache. For both data blocks and tags, we observe that only a small portion of cache entries (data and tags) experience the highest write ratio (>90% relative to the entry with the maximum number of writes) and, as a result, they indicate severe RMS-EM conditions for all cache entries. We observe that the ratio of the average number of writes per entry to the maximum number of writes of all entries is approximately 50%. A similar result for write operations on cache lines was also obtained by Valero et al. in their study of the different aspects of cache reliability [19].

Examination of Figure 6 and Figure 7 shows that the benchmark 649.fotonik3d, behaves differently than all other benchmarks. This is explained by the fact that 649.fotonik3d has write distribution that are spread uniformly over most cache lines.

Figure 8 shows a histogram for L3 writes for cache data lines. For most benchmarks, the number of writes is very small for the majority of cache data lines, where almost all of them experience 25% or less write operations relative to a very small portion of cache lines with the maximum number of writes. Overall, the ratio of the average number of write operations to the maximum number of writes is 8%.

Our experiments, which also include an analysis of cache tag writes, indicate that tag writes spread more uniformly in compare data lines, and the majority of cache tags experience smaller variance in the number of writes. The ratio of the average number of tag writes to the maximum number of tag writes is nearly 70% on average for the L1-D cache and approximately 50% for L2 and L3 tags.

The results presented in this section, support our observation that cache data lines experience a switching probability distribution with high variance and with a minority of lines being highly stressed by the maximum number of write operations and, as a result, dictate, much more severe RMS-EM conditions for the entire cache. Similar conclusions are obtained from our observation of registers write access and ALU use where, in both cases, the switching probability induced by the workload is nonuniformly distributed. Such behavior leads to an over-design condition for RMS-EM that can degrade overall performance and increase IC area. In the next section, we propose EM-aware microarchitectural mechanisms to smooth the switching probability hotspots and thereby mitigate RMS-EM reliability impact. This approach results in a dramatic relaxation of the overall RMS-EM sign-off design conditions.

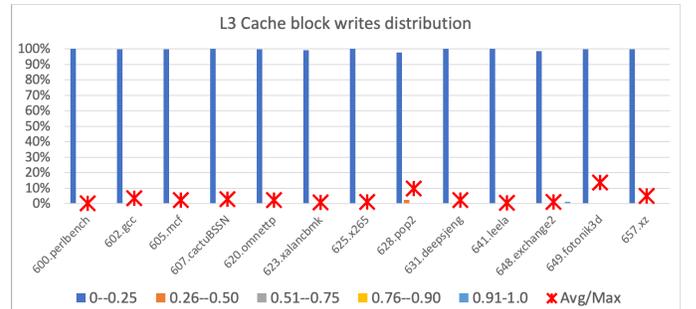

*Figure 8 - Distribution of L3 cache block writes*

## IV. Proposed RMS-Electromigration-Aware Resource-Allocation Mechanism

This section introduces our architecture solutions to eliminate switching probability hotspots and thereby relax RMS-EM sign-off conditions. The principle of our proposed solutions is similar to those employed in the area of workload balancing in computer systems. The idea is based on switching probability aware resource allocation scheme that smoothens the utilization of the available computational resources uniformly. As a result, RMS-EM reliability impact is significantly reduced. The following subsections introduce RMS EM-aware architectures for dealing with RMS-EM switching probability related hotspots on ALU execution units, register files, and cache memories, respectively. The novelty of the proposed solutions may be summarized as follows:

1. We offer RMS EM-aware architectural solutions dedicated to fundamental microprocessor building elements: register files, execution units and cache memories, whereas prior studies made limited use of such information.

2. The proposed solutions can be incorporated in conjunction with physical-design-based flows and provide a complementary enhancement to such flows.

3. We avoid the need to duplicate logic, reduce performance, or employ dedicated mechanisms to detect EM degradation through normal IC operation that were suggested by [6].

4. The proposed solution eliminates the dynamic power overhead and the design complexity suggested by past studies such as [8].

5. Finally, we avoid compromising on reliability and management, as suggested by Refs. [5,20].

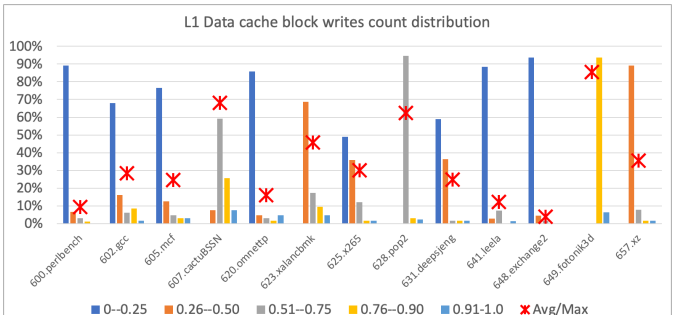

*Figure 6 - Distribution of L1-D cache block writes*

As part of introducing the principles of our solutions we also summarize the limitations of the proposed techniques to the following cases:

1. Our study is limited to digital circuits. Analog circuits are out of the scope of this study.

2. Our solutions are highly effective when the switching probability is a dominant factor in inducing RMS-EM. System with low activity rate may have limited benefit from the proposed techniques.

3. Our solutions rely on non-uniform distribution of the switching probability that can be exploited to smoothen RMS-EM hotspots. When the switching probability is evenly distributed the effectiveness of out techniques are limited.

## A. Electromigration-Aware ALU Allocation

In the previous section, we observed that ALUs are not utilized in an RMS EM-aware manner, which means that the maximum switching probability is dictated by a small, over-used subset of ALUs. The proposed RMS EM-aware scheme assumes that all pending ALU instructions are allocated to a centralized instruction queue, and in each cycle a scheduler allocates ALUs to execution-ready instructions. Although the proposed scheme is described for ALUs, it can also be applied to any type of multi-execution unit employed by microprocessors.

In this study, we present two alternatives that implement the same basic principle in different ways. The aim of both solutions is to start allocating the resources from a different leading point each time. The first simple solution is to have a counter (e.g. 32-bit counter) that is incremented each clock cycle and wraps around when expired so that the leading resource number to use is calculated as counter value modulo the number of physical resources. Thus, for our simulated environment, we assume N = 3. When the counter expires, we reset its content and continue with the allocation in the next cycle.

The second solution is illustrated in Algorithm 1; here, we extend each resource with a single bit (Ex_counter) and add a single global bit (Global_counter) for the overall management of the allocation. All counters are initialized to zero. We suggest that the EM-aware allocation algorithm selects execution units whose corresponding counter state equals the global counter (denoted by the set M). If the number of available execution units that satisfy this condition exceeds the required number of instructions to be issued (k< |M|), then a subset, Q⊂M, (based on the required number of instructions to be issued) of those execution units is selected, and all their corresponding counters are switched (between zero and one). Otherwise, the set M of all execution units with their counter state equal to the global counter are selected while the rest of the execution units needed to satisfy the required instruction to be issued are selected from the set of other pool of ALUs, Q ⊆ U\M (such that |Q|= k-|M|), whose counter is not equal to the global counter. In this case, only the global counter and the Ex_counters which are equal to the global counter are incremented.

---

**Algorithm 1** – EM-aware execution-unit allocation:

**Input:** k<N number of execution units to be allocated.
**Output:** Vector E= $(e_0, e_1, \ldots, e_{n-1})$, for every $0 \le i \le n-1$, only if $e_i$=1 execution unit i to be allocated, otherwise not allocated.
**Initialization:** Ex_counter[i]=0 for every $0 \le i \le n-1$, Global_counter=0
1. M = {$0 \le i \le n-1$ | Ex_counter[i]= Global_counter}
**2. if** k< |M| **then**
3.    let Q⊂M such that |Q|= k
4.    $e_i$=1 for every i∈Q, otherwise $e_i$=0
5.    Ex_counter[i]++ for every i∈Q
**6. end if**
**7. else** // k≥ |M|
8.    let Q ⊆ U\M such that |Q|= k-|M|
9.    $e_i$=1 for every i∈Q∪M, otherwise $e_i$=0
10.   Ex_counter[i]++ for every i∈ Q∪M
**11.   Global_counter++**
**12. end else**
**13. return** E

---

Table 2 shows an example of the algorithm output for three ALUs.

*Table 2 - Example of EM-aware ALU scheduling*

| Clock cycle | Issued instructions | Ex_counter[2:0] | Global counter | Selected ALU(s) |
|---|---|---|---|---|
| 0 | 0 | 0, 0, 0 | 0 | None |
| 1 | 2 | 0, 1, 1 | 0 | 0, 1 |
| 2 | 2 | 1, 1, 0 | 1 | 2, 0 |
| 3 | 3 | 0, 0, 1 | 0 | 1, 2, 0 |

The implementation of the first solution is straightforward and may perform well given a large number of execution units. The implementation of the second solution is more complicated, but our implementation trial indicates that it can be done with negligible overhead in terms of logical area and computation time for both the ALU-selection logic and the counter-incrementation logic. The following table summarizes power, timing, and area overhead for a 28 nm process. It should be noted that the proposed solution does not affect timing since the counters are updated in parallel to the ALU execution cycle. In addition, we compare the routing resources used by the two options and find that both use negligible routing resources. Option 1 uses 50 nets with a total wire length of 51 µm using M1-M4 metal layers. Option 2 uses 57 nets with a total wire length of 299 µm using M1-M5 metal layers. Note that the total net length of the original design is 21,255um and therefore in both options the wire length overhead is relatively negligible (0.23% and 0.14% for option 1 and 2 respectively).

*Table 3 - ALU scheduling overhead*

| Option | Original Area [um²] | Area Overhead [um²]/[%] | Original Power [uW] | Power [uW]/ [%] | Timing impact |
|--------|---------|---------|---------|---------|---------|
| 1 | 200613 | 316 / 0.15 | 641.79 | 0.031 / 0.004% | None (reg-to-reg delay < clock cycle time) |
| 2 | 200613 | 85.9 / 0.04 | 641.79 | 0.026 / 0.004% | None (reg-to-reg delay < clock cycle time) |

### B. Electromigration-Aware Registers Allocation

The results of the measurements presented in Section 3 clearly indicate that write operations to registers are not uniformly distributed. Moreover, specific registers (e.g., RAX) experienced an excessive number of writes. Such behavior by a small number of registers dictates difficult RMS-EM conditions for all registers and may result in reliability concerns. Note that this section deals mainly with architectural registers assigned by the compiler rather than with physical registers implemented by the out-of-order (OoO) microprocessors. For the latter, physical registers are usually implemented as a cyclic buffer within the reorder buffer and, as a result, all writes are spread uniformly over time.

The proposed architectural solution, illustrated in Figure 9, avoids write hotspots in registers by periodically changing the mapping of registers to their corresponding architectural hosting locations. The scheme is based on modulo rotation of the mapping between the architectural register identifier and its physical locations. As illustrated in Figure 9, a pulse trigger is asserted to shift the register mapping in the register-file (RF) either periodically (or each time we change CR3) or as part of the return-from-interrupt procedure before saving the values of the user-level process. A modulo-counter (RF rotator) serves to map the architectural register number to the physical register location. The physical location of each architectural register is determined by summing the architectural register identifier with the RF rotator value. In addition, 2-to-1 multiplexors are inserted between adjacent registers to select between the functional RF write-port and the value in the adjacent register (which is selected by a trigger assertion). After each assertion of the rotation trigger (at any arbitrary time point), the counter is incremented, and the physical register point are shifted between neighboring registers by changing the control of the multiplexors and asserting the load-enable control signal of the registers.

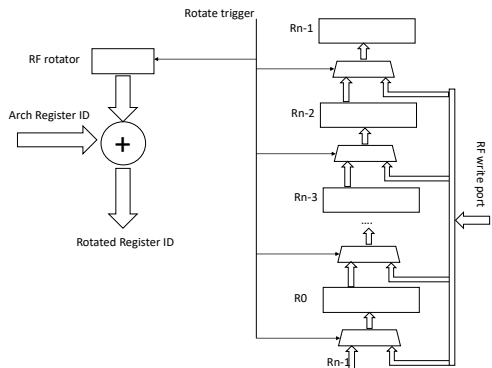

*Figure 9 - Scheme for electromigration-aware RF mapping*

The following table summarize power, timing path and area overhead for 28nm process (for 32 GPRs):

*Table 4 - GPR rotation overhead*

| Original Area [um²] | Area Overhead [um²]/[%] | Original Power [uW] | Power [uW] /[%] | Timing impact |
|--------|---------|---------|---------|---------|
| 77234 | 1973 / 2.5 | 20,162 | 0.282 / 0.001% | 50ps delay added to access time |

The proposed solution has a certain similarity to the Sun SPARC and Berkley RISC CPUs register window [26], which is used for different purposes. Register window is a scheme that aims to evenly distribute sets of GPR registers between different sections of code, typically procedure calls, and upon every nested call the register window is shifted to provide the program a new working set of registers. Unlike, the register window technique which is limited to integer registers our proposed scheme is extended to all architectural registers (FP, vector, control etc.). Note that the register window involves more frequent register-window switching, resulting in excessive dynamic power whereas the rotation frequency of our proposed scheme is very low.

### C. Electromigration-Aware Cache Memories

Cache memories may generate RMS-EM hot spots in various cache lines that are nonuniformly. Note that, in this subsection, the term "cache" refers to any architectural structure that uses a cache organization (e.g., TLBs). As a result, a small fraction of cache lines dictates the worst RMS-EM scenario for the entire cache. The principal of the proposed EM-aware cache memory scheme, illustrated in Figure 10, is based on similar principles of the register file solution.

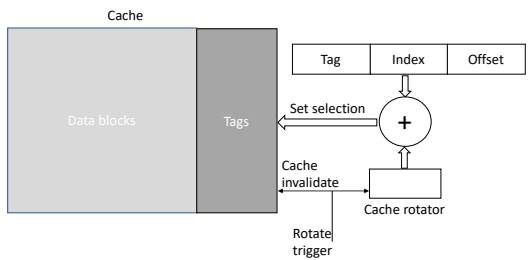

*Figure 10 - Electromigration-aware cache memory mapping*

Our proposed scheme is similar to two previously introduced solutions for different physical reliability problems. The first one, introduced by Calimera et al. [27] to handle a different problem related to SRAM asymmetric aging, suggested re-indexing cache lines using various mapping functions. The second method was introduced by Wang et al. [29] to mitigate write endurance in PCM-based non-volatile memories (NVMs). PCM-based NVMs experience bitcell wearout after an excessive number of writes leaving the bitcell resistance in a low- or high-resistance state. This may happen due to either Ge depletion in the bitcell area or when the heating electrode is detached [30]. Their proposed technique (similar to Calimera et al.) suggested Swap-Shift method to swap a pair of cache sets whenever the number of writes reaches a certain threshold.

By using a similar technique, our proposed scheme avoids hotspots of cache writes by periodically changing the cache set mapping of cache blocks to their corresponding physical cache lines. As with the RF solution, the principal of this scheme is based on modulo rotation of the mapping between the set field (taken from the block address) and its physical set location. Our suggested mapping method calculates the cache set number by adding the block index field to a modulo-counter. The counter determines the physical index shifting relative to the cache block original index. A periodic pulse trigger is maintained to shift the mapping of the cache sets. After each assertion of the pulse trigger, the modulo counter is incremented, and all cache lines are invalidated. Note that this cache-invalidation circuitry already exists in many modern microprocessors for the purpose of cache-context invalidation, so the proposed method does not incur an overhead by adding this mechanism. In addition, the periodic pulse trigger can operate at a relatively low frequency to ensure negligible performance overhead due to the cache invalidation. The end result is that this approach avoids write hotspots by periodically spreading the cache lines mapping across all cache sets. The following table summarizes the power, critical timing path impact and area overhead for the 28 nm process:

*Table 5 - Cache rotator overhead*

| Cache index size | Orig. Area [mm²] | Area Overhead [um²]/[%] | Orig. Power [mW] | Power Overhead [nW] /[%] | Timing impact Delay added to access time [ps] |
|---|---|---|---|---|---|
| 6 bits (L1-D) | 3.07 | 104/ 0.003% | 481 | 8/ 0.000% | 60 |
| 7 bits (L1-I) | 2.99 | 123/ 0.004% | 480 | 10.2/ 0.000% | 63 |
| 9 bits (L2) | 8.07 | 157/ 0.002% | 818 | 15.6/ 0.000% | 67 |
| 13 bits (L3) | 48.21 | 226/ 0.000% | 8536 | 29.5/ 0.000% | 76 |

To avoid the potential overhead incurred by flushing the cache content (and by the write-back of all the dirty lines), we suggest doing the operation either very infrequently or by exploiting events that require flushing these structures (e.g., after a sleep mode when all caches were cleaned).

## V. Experimental Study of RMS Electromigration-Aware Architecture

This section presents the experimental results for the proposed architecture solutions (presented in the previous section) to reduce the impact of RMS-EM. The metric of MTF improvement is defined as the increase in the RMS EM-aware MTF with respect to the original MTF and by applying Equation 4 the following equation is obtained:

$$MTF\ improvement = \frac{MTF_{RMS\ EM-aware}}{MTF_{original}} - 1$$
$$= \frac{p_{max\ original}}{p_{max\ RMS\ EM-aware}} - 1$$

*Equation 7 - MTF Improvement*

where $p_{max\ RMS\ EM-aware}$ and $p_{max\ original}$ are the maximum toggle rates of a module with RMS EM-aware architecture and the original architecture respectively.

Note that our proposed techniques in Section 4 did not report performance overhead, so this section focuses on how the algorithms proposed herein affect the MTF improvement. Our experimental analysis starts by examining the improvement of RMS-EM MTF provided by the proposed solution through relaxing the maximum switching probability. Next, we validate our experimental observation via extensive physical RMS-EM simulations that consider the Joule-heating effect through dynamic high-resolution thermal analyses.

### A. Toggle rate based experimental analysis for RMS-EM MTF improvement

We first examine an RMS EM-aware solution for ALU execution units. Figure 11 shows how the solution describes in Algorithm 1 affects the RMS-EM MTF for the SPEC2017 benchmarks. Examination of the two solutions introduced in previous section indicates that they behave very similarly. The results show that the proposed algorithm efficiently eliminates

ALU usage hotspots and can potentially improve RMS-EM MTF by approximately 100% on average. The results vary from nearly 34% potential MTF improvement up to 130% improvement. This result is because the proposed scheme distributes ALU use uniformly and reduces the RMS-EM hotspots.

As part of this study, we also compare the instructions per cycle (IPC) versus the potential RMS-EM MTF improvement, as shown in Figure 12. Benchmarks with small IPC have a greater potential for RMS-EM MTF improvement because of the underused ALUs that could potentially help reduce the maximum RMS-EM hotspots.

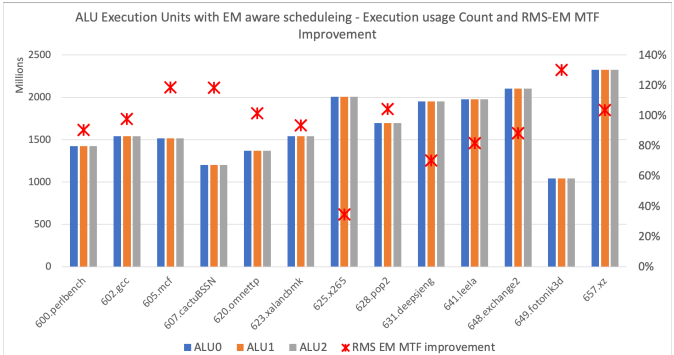

*Figure 11 - Distribution of ALU execution usage count and MTF improvement with RMS electromigration-aware allocation*

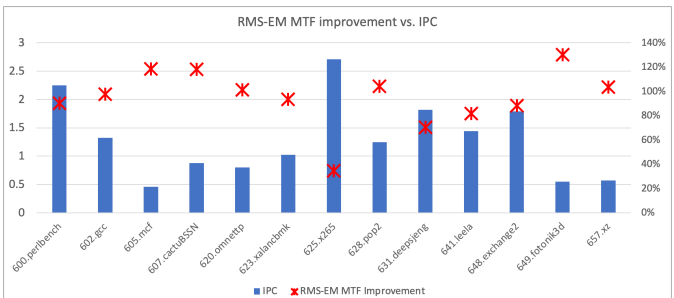

*Figure 12 - ALU electromigration stress reduction vs. IPC*

The next results show the potential RMS-EM MTF improvement obtained by the proposed architectural solution for both the GPR register file and FP register file (Figure 13 and Figure 14, respectively). For both register files, the number of writes is distributed uniformly over all registers, and no hotspots exist. In addition, the MTF potentially improves by nearly 400% on average for the GPR registers and 1200% on average for the FP registers. The rotation trigger in the simulation was asserted every 10 million clock cycles. We examined different rotation trigger rates and found that this value does not impact performance.

As part of the experiments, we also observed that the flags and stack-pointer registers experienced excessive stress of write operations, which makes them highly susceptible to RMS-EM. Figure 15 illustrates the number of write operations to the flags register and stack-pointer register and compares them with the maximum number of writes per register in the GPR register file. For almost all benchmarks, the number of writes to the flags register significantly exceeds those to the GPR and stack-pointer registers. This result is due to the fact that almost every instruction involves implicit writes to the flag register, which motivates us to extend the EM-aware scheme proposed for the GPR register file to include both the flag and stack-pointer registers. Figure 15 shows that, in this case, the maximum number of write operations is reduced even more (varying from 80% to >90%) and that the potential MTF improvement is over 760% on average.

The last part of this section is devoted to examining the RMS-EM MTF improvement for the TLBs and cache data lines. The results are illustrated in Figure 16 show that in most cases, the RMS-EM stress is significantly reduced as a result of the repetitive rotation of the set mapping and invalidation. This helps to distribute write operations uniformly over all sets and ways. For the D-TLB, we suggest triggering the rotation either when the TLB is flushed by the system, or by performing a period rotation (e.g., every 10M TLB accesses). For the L1-D cache, we suggest a similar periodic rotation trigger every 10M accesses. For all these options, the performance overhead is minimal. As previously discussed, for both L2 and L3, we suggest triggering the set rotation upon each system wakeup from sleep mode. In this case, no performance overhead is incurred. In our simulation we use an interval of 10M cache accesses, the same trigger duration of the L1-D cache for both the L2 and L3 caches.

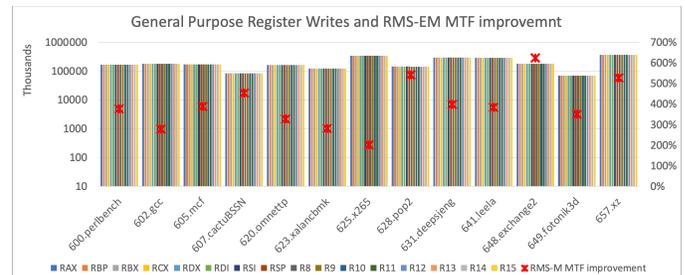

*Figure 13 - GPR writes distribution with RMS-EM MTF improvement*

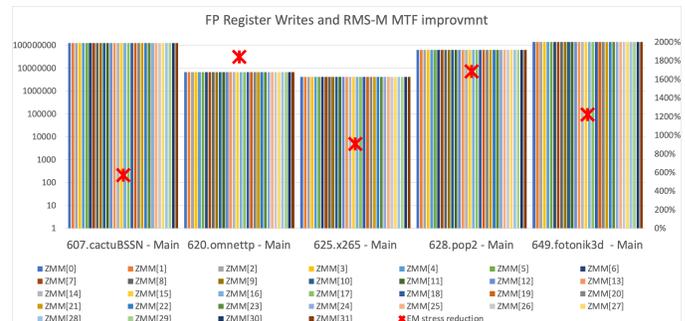

*Figure 14 - FP register writes distribution with RMS-EM MTF improvement*

Figure 16 illustrates the RMS-EM MTF improvement for DTLB, L1-D, L2, and L3 caches. The geometric mean MTF improvement for the DTLB, L1-D, L2 and L3 caches is 65%, 230%, 86%, 4670% respectively. Note that the experimental results of the EM-aware architectural solution are consistent with the results presented in Section 3. These figures suggest that a smaller ratio of the average number of write operations to the maximum number of writes corresponds to greater RMS-EM MTF improvement.

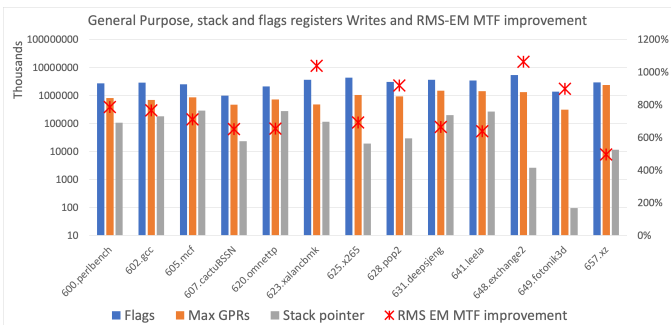

*Figure 15 - Distribution of GPRs, flags, and stack pointer writes with EM-aware allocation*

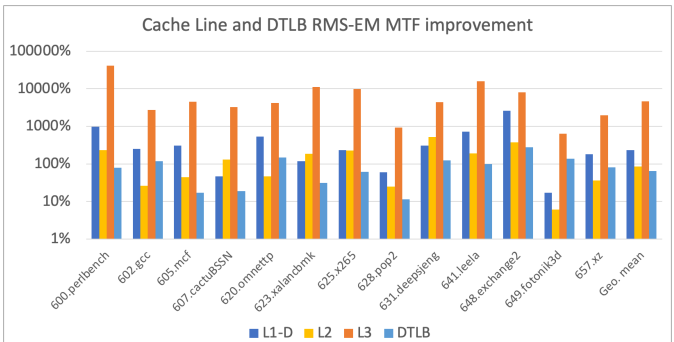

*Figure 16 - Cache lines electromigration RMS-EM MTF Improvement*

Based on the experimental results, we observe that RMS-EM MTF can be significantly extended in the microprocessors building blocks that are examined. The observations detailed herein reveal an average improvement in RMS-EM MTF by 100% for ALUs, 400% for the integer register files, 1200% for FP register file and 86%–4670% for cache data blocks. These results indicate that the proposed EM-aware solution should allow microprocessor designers to significantly relax the maximum switching probability and, as a result, to avoid a significant number of potential RMS-EM violations.

Alternatively, the reduction in the maximum switching rate translates into an extended device lifetime. Since the RMS-EM MTF and device lifetime depend not only on the switching probability but also on the electrical and thermal characteristics of the circuit, we extend our experimental analysis by performing physical simulations that consider the Joule-heating effect and toggle rate.

### B. Physical RMS-EM simulations based on Joule-heating effect

In the last part of our experimental analysis, we present extensive physical simulations that consider both the toggle rate and the Joule-heating effect through a dynamic, high-resolution thermal analysis. The simulations were implemented in the Cadence® Voltus™ simulation environment [31], which performs detailed RMS EM analysis of the Joule-heating effect and self-heating under different toggle rates. Voltus™ is considered an industry standard for EM sign-off and is certified as an EM sign-off tool by many foundries. The simulation environment takes into account the parameters of transistors that contribute to the RMS current, such as drive strength (fins, number of fingers), channel length, and channel width. The tool makes detailed RMS current calculations to analyze the Joule-heating effect and self-heating on all signal wires while taking into account metal dimensions and type (power-grid connections are beyond the scope of this analysis). As part of the simulation process, the tool also certifies that the calculated RMS current of every net does not exceed the maximum RMS current, which is considered a mandatory reliability criterion and is specified in the foundry technology file ([37]).

The Voltus™ environment requires to synthesize and place-and-route the design under test. The full implementation flow was done on the three architectural structures introduced in Section IV: ALU, Register-files, and L1 data cache memory. The design parameters, implementation tools, and simulation environment are summarized in Table 6. Through the RMS-EM simulations, Voltus™ calculates the RMS current, $I_{RMS}$, for every metal net in the design while considering the toggle rate obtained from the functional simulations. The technology file, which is provided by the foundry, specifies the maximum allowed RMS current, $I_{RMS\_MAX}$, per metal layer based on its physical properties (physical dimension and material type).

*Table 6 – Implementation and RMS-EM simulation tools and design parameters*

| Implementation and EM simulation tools and design parameters | |
|---|---|
| Synthesis tool | Cadence® Genus™ version 19.11-s087_1 |
| Place-and-route tool | Cadence® Innovus™ version 19.11-s128_1 |
| EM tool | Cadence® Voltus™ version 19.11-s129_1 |
| Process | 28 nm |
| Clock frequency | 2.66 GHz |
| Core voltage | 0.9 V |
| Tj | 105°C (self-heating is modeled by the Voltus™ simulation environment) |
| Metal layers | Metal 1 to metal 9 |

Figure 17 summarizes the reduction of the ratio of $I_{RMS}$ to $I_{RMS\_MAX}$ in the design with the EM-aware architecture versus the original design for each benchmark that we use. In addition, it presents the percentage of metal nets that can leverage such RMS current reduction. The results show that, by taking the average over all benchmarks, 62% of cache memory nets can leverage a 36% reduction in their RMS current. For the ALU, nearly all nets can leverage a reduction of approximately 30% in RMS current while 55% of the RF nets experience 68% reduction in their RMS current. Note that the metal nets that do not leverage a reduction in the RMS current, already exhibited very small RMS current, so their overall improvement is not noticeable by the tool. The extended MTF as a result of $I_{RMS}$ reduction can be calculated using Equation 3. The extended MTF is proportional to the ratio of $I_{RMS\_MAX}$ to the reduced $I_{RMS}$ to the power of two. Thus, the observed RMS current reduction offers at least x2, x10, x2.5 lifetime extension for ALUs, register files, and cache memories, respectively. One may note that the extended MTF experimental results which are obtained using the RMS-EM physical simulation are similar to the MTF improvement prediction provided by the experimental results provided by Figure 16 which were based on the switching probability reduction.

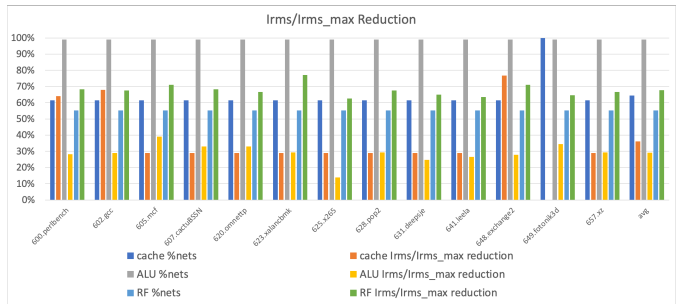

*Figure 17 – Physical RMS EM simulations: $I_{RMS}/I_{RMS\text{-}Max}$ ratio reduction of the EM-aware architecture with respect to the original design measurements*

## VI. CONCLUSIONS

Microprocessor reliability is a crucial requirement that introduces major micro-architectural and design challenges. Traditionally, reliability and RMS-EM related issues are handled at the physical design level that enforces design rules using worst case scenario analysis in order to detect violations and attempts to solve them. In our study we presented RMS EM-aware micro-architectural solution that can significantly relax the over-design of traditional methods and significantly extend microprocessor lifetime.

This paper indicates that microprocessors are highly susceptible to RMS-EM because they process highly variable dynamic workloads on non-EM-aware microarchitectures. We introduce herein architectural solutions that take into account the RMS-EM effect and reduce excess use of execution units and write operations to registers and memory-hierarchy elements. The principal of the proposed solutions is based on RMS EM-aware resource allocation that attempts to uniformly distribute write operations and the use of computational elements over all available resources. This solution can be incorporated into physical-design-based approaches where it offers a complementary enhancement to existing methods. Our analysis shows that the proposed solutions incur minor area and power overhead and negligible performance degradation with respect to prior studies. In addition, our experimental results indicate that the proposed architecture significantly relaxes the RMS-EM switching probability sign-off conditions by 50% for ALUs, 80%–90% for the register files, and 46%–92% for the data blocks of cache memories. Our RMS EM physical simulations indicate that such toggle rate relaxation leads to a dramatic reduction in $I_{RMS}$ of 30%, 68%, and 36% for ALUs, register files, and cache memories respectively. Such a reduction translates into lifetime extension of at least x2, x10, x2.5 for ALUs, register files, and cache memories respectively.

EM has become a major challenge in advanced technologies, and further studies are required to continue exploring new architectures and to identify other avenues to reduce EM and extend device lifetime. In this study, we examined how RMS-EM affects modern microprocessors, although the approach used herein may be extended to other processing elements such as security engines, GPUs, and TPUs. We also encourage future studies to examine software-based solutions for RMS-EM reduction.

## REFERENCES

[1] X. Xuan, Analysis and Design of Reliable Mixed-Signal CMOS Circuits, PhD thesis, Georgia Inst. of Technology, Dept. of Electrical and Computer Engineering, 2004.

[2] J. Lienig and G. Jerke, Embedded Tutorial: Electromigration-Aware Physical Design of Integrated Circuits, Proc. 18th Int'l Conf. VLSI Design (VLSID 05), IEEE Press, 2005, pp. 77-82.

[3] J. Lienig, Introduction to electromigration-aware physical design. In Proceedings of the International Symposium on Physical Design (ISPD'06). ACM, New York, 39–46.

[4] J. Lienig. Electromigration and Its Impact on Physical Design in Future Technologies. Proceedings of the 2013 ACM International symposium on Physical Design, March 2013.

[5] J. Srinivasan, S. V. Adve, P. Bose and J. A. Rivers. Lifetime Reliability: Toward an Architectural Solution. IEEE Micro, special issue on Emerging Trends, vol. 25, issue 3, May-June 2005, 2-12.

[6] J. Srinivasan, S.V. Adve, P. Bose and J. A. Rivers. Exploiting Structural Duplication for Lifetime Reliability Enhancement. Proceed. of the 32nd International Symposium on Computer Architecture June 2005.

[7] A. Dasgupta and R. Karri, Electromigration Reliability Enhancement Via Bus Activity Distribution, Proc. 33rd Ann. Conf. Design Automation (DAC 96), ACM Press, 1996, pp. 353-356.

[8] J. Abella, X. Vera, O S. U. O. Ergin, A. González and J. W. Tschanz. Refueling: Preventing Wire Degradation due to Electromigration. IEEE Micro (Volume: 28 , Issue: 6 , Nov.-Dec. 2008 ).

[9] J. Tao, J. Chen, N. Cheung, and C. Hu. (1996). Modeling and characterization of electromigration failures under bidirectional current stress. IEEE Transactions on Electron Devices, 43, 800-808.

[10] J. Abella and X. Vera, Electromigration for Microarchitects. ACM Computing Surveys (CSUR)March 2010 Article No.: 9

[11] Operating Temperature, Wikipedia - https://en.wikipedia.org/wiki/Operating_temperature.

[12] Failure Mechanism based Stress test Qualification for Integrated Circuit. Automotive Electronics Council, Component Technical Committee - AEC - Q100 - REV-G standard.

[13] J. R. Black, "Electromigration – A brief survey and some recent results," IEEE Trans. on Electronic Devices (April 1969), 338-347. DOI= http://dx.doi.org/10.1109/T-ED.1969.16754

[14] A. B. Kahng, S. Nath and T. S. Rosing, On Potential Design Impacts of Electromigration Awareness. 2013 18th Asia and South Pacific Design Automation Conference (ASP-DAC)

[15] I. A. Blech, Electromigration in thin aluminum films on titanium nitride, J. Appl. Phys., vol. 47 (1976), 1203–1208. http://dx.doi.org/10.1063/1.322842

[16] C. S. Hau-Riege, An introduction to Cu electromigration, Microel. Reliab., vol. 44 (2004), 195–205. DOI= http://dx.doi.org/10.1016/j.microrel.2003.10.020

[17] A. Scorzoni, B. Neri, C. Caprile, F. Fantini, Electromigration in thin- film inter-connection lines: models, methods and results, Material Science Reports, New York: Elsevier, vol. 7 (1991), 143–219. http://dx.doi.org/10.1016/0920-2307(91)90005-8

[18] D. Young, A. Christou, Failure mechanism models for electromigration, IEEE Trans. on Reliability, vol. 43(2) (June 1994), 186–192. DOI= http://dx.doi.org/10.1109/24.294984

[19] A. Valero, N. Miralaei, S. Petit, J. Sahuquillo, and T. M. Jones. On Microarchitectural Mechanisms for Cache Wearout Reduction. IEEE Transactions on Very Large-Scale Integration (VLSI) Systems, Vol. 25, No. 3, March 2017.

[20] J. Srinivasan, S.V. Adve, P. Bose and J. A. Rivers, The Case for Lifetime Reliability-Aware Microprocessors, Proceedings of 31st International Symposium on Computer Architecture (ISCA '04) June 2004.

[21] T. E. Carlson, W. Heirman, and L. Eeckhout. Sniper: Exploring the level of abstraction for scalable and accurate parallel multi-core simulations. In Proceedings of the International Conference for High Performance Computing, Net- working, Storage and Analysis (SC), Nov. 2011.

[22] M. E. Thomadakis. The architecture of the Nehalem processor and Nehalem-EP smp platforms. Technical report, December 2010. http://sc.tamu.edu/systems/eos/nehalem.pdf.

[23] A. Limaye and T. Adegbija, "A workload characterization of the spec cpu2017 benchmark suite," in 2018 IEEE International Symposium on

Performance Analysis of Systems and Software (ISPASS), pp. 149–158, April 2018

[24] Q. Wu, S. Flolid, S. Song, J. Deng, L. K. John. Hot Regions in SPEC CPU2017. 2018 IEEE International Symposium on Workload Characterization (IISWC).

[25] K-T Jang, Y-J Park, MW Jeong, S-M Lim, H-W Yeon, J-Y Cho, M-G Jin, J-S Shin, B-W Woo, J-Y Bae, Y-C Hwang, Y-C Joo. Electromigration behavior of advanced metallization on the structural effects for memory devices. Microelectronic Engineering Vol. 156, 20 April 2016.

[26] The SPARC Architecture Manual, Version 8.

[27] A. Calimera, M. Loghi, E. Macii and M. Poncino, "Dynamic Indexing: Leakage-Aging Co-Optimization for Caches," in IEEE Transactions on Computer-Aided Design of Integrated Circuits and Systems, vol. 33, no. 2, pp. 251-264, Feb. 2014, doi: 10.1109/TCAD.2013.2287187

[28] K. Swaminathan, N. Chandramoorthy, C. Cher, R. Bertran, A. Buyuktosunoglu and P. Bose, "BRAVO: Balanced Reliability-Aware Voltage Optimization," 2017 IEEE International Symposium on High Performance Computer Architecture (HPCA), Austin, TX, 2017, pp. 97-108, doi: 10.1109/HPCA.2017.56.

[29] J. Wang, X. Dong, Y. Xie and N. P. Jouppi, "i2WAP: Improving non-volatile cache lifetime by reducing inter- and intra-set write variations," 2013 IEEE 19th International Symposium on High Performance Computer Architecture (HPCA), Shenzhen, 2013, pp. 234-245, doi: 10.1109/HPCA.2013.6522322.

[30] M. K. Tavana, A. K. Ziabari, M. Arjomand, M. Kandemir, C. Das, and D. Kaeli. 2017. REMAP: a reliability/endurance mechanism for advancing PCM. In Proceedings of the International Symposium on Memory Systems (MEMSYS '17). Association for Computing Machinery, New York, NY, USA, 385–398. DOI:https://doi.org/10.1145/3132402.3132421

[31] Voltus$^{TM}$ User Guide. http://www.cadence.com

[32] J. A. Maiz, Characterization of electromigration under bidirectional (BC) and pulsed unidirectional (PDC) currents," in Proc. Int. Reliab. Phys. Conf. (IRPS), 1989, pp. 220–228.

[33] K. Jonggook, V. C. Tyree, C. R. Crowell, Temperature gradient effects in electromigration using an extended transition probability model and temperature gradient free tests. I. Transition probability model, *IEEE Int. Integrated Reliability Workshop Final Report* (1999), 24 -40. DOI= http://dx.doi.org/10.1109/IRWS.1999.830555

[34] X. Yu, K. Weide, A study of the thermal-electrical- and mechanical influence on degradation in an aluminum-pad structure, *Microelectronics and Reliability* (1997), 37, 1545 – 1548. DOI= http://dx.doi.org/10.1016/S0026-2714(97)00105-4

[35] Sheldon X.-D. Tan, Mehdi Tahoori, Taeyoung Kim, Shengcheng Wang, Zeyu Sun and Saman Kiamehr, "VLSI Systems Long-Term Reliability -- Modeling, Simulation and Optimization", Springer Publisher, 2019. DOI: 10.1007/978-3-030-26172-6, ISBN: 978-3-030-26171-9

[36] R. Monig, R.R. Keller, C.A. Volkert, Thermal fatigue testing of thin metal films. Review of Scientific Instruments 75(11), 4997–5004 (2004)

[37] LEF DEF reference. http://www.si2.org/openeda.si2.org/

[38] S. Wang, T. Kim, Z. Sun, S. X.-D. Tan, M. Tahoori, "Recovery-aware proactive TSV repair for electromigration lifetime enhancement in 3D ICs", IEEE Transactions on Very Large Scale Integrated Systems (TVLSI), Vol. 26, no. 3, pp. 531-543, March 2018. DOI: 10.1109/TVLSI.2017.2775586

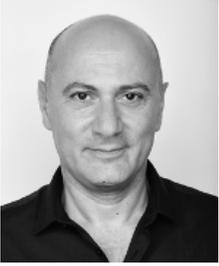 **Freddy Gabbay** is an Associate Professor and Head of Computer Science Department at the Ruppin Academic Center, Israel and Adjunct Associate Professor at the Technion Israel Institute of technology. He graduated his B.Sc., M.Sc. and Ph.D. at the EE department of the Technion - Israel Institute of Technology, Haifa, Israel in 1994, 1995 and 1998 respectively. His areas of research are HPC accelerators, VLSI design, chip reliability, microprocessor architecture and machine learning.

In 1998, he worked as a researcher at Intel Micro-processor Research Lab. In 1999 he joined Mellanox Technologies and held various positions in leading various product lines architecture and ASIC design. In 2003, he joined Freescale Semiconductor as a senior design manager and led the baseband ASIC products. In 2012 he joined back Mellanox Technologies where he served as Vice President of Chip Design. Prof. Gabbay also holds 19 patents and is an IEEE member.

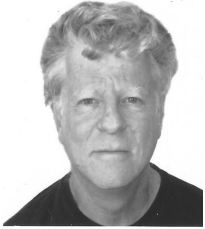 **Avi Mendelson** is a visiting professor at the CS and EE departments at the Technion, Israel Institute of Technology, Haifa, Israel and in the EE department, NTU, Singapore. He has a blend of industrial and academic experience in several different areas such as Computer architecture, Power management, security, and Real-Time Systems.

As part of his industrial role, he worked for National semiconductor, in the team that invented and developed the first PC-on-the-Chip. At Intel he worked 5 years as a researcher in Intel research labs and 6 years as principle engineer in the mobile CPU architecture team where he was chief architecture of the first CMP feature (multicore) of Intel cores. For this task and leadership, he got the IAA (Intel Achievement Award)

Prof. Avi Mendelson is an IEEE Fellow, was a member of the Board of Governors of the IEEE Computer Society and served as a second VP of the IEEE Computer Society.